# Modeling electricity demand, welfare function and elasticity of electricity demand based on the customers' risk aversion behavior

Amir Niromandfam[1,2*], Saeid Pashaei Choboghloo[3]
[1] Faculty of Electrical Engineering, Renewable Energy Research Center, Sahand University of Technology, Tabriz, Iran
[2] Daneshmand Research and Development Institute, Tehran, Iran
[3] Faculty of Electrical Engineering, University of Zanjan, Zanjan, Iran
a_niromandfam@sut.ac.ir[*], saeidpashaei.ch@gmail.com

**Abstract:** Nowadays, attitudes towards electricity customers have been changed, so that they are no longer considered static players. The customers' behavior identification is vital for establishing modern power systems. This paper utilizes the customers' risk aversion coefficient to model their consumption behavior. Through the coefficients, the value of electricity consumption is measured employing the concept of the utility function. The utility function is an essential concept in microeconomics that measures customers' preferences and interprets how a rational consumer would make consumption decisions. Based on the utility function, an economic load model and the customers' welfare function are formulated. The load model is utilized to estimate price elasticity, and income elasticity of electricity demand. The results illustrated as the risk aversion coefficient increases, the consumers achieve more satisfaction from the consumption. Electricity consumption will be complement (substitute) if the risk aversion coefficient is greater (lower) than the unit. For the unit risk aversion coefficient, the cross-price elasticity will be zero, which means the consumption is independent of the other period demand. The results illustrated electricity price increment neutralizes the income increment effect on consumption. Hence, to study electricity consumption, the price and income variations should be considered, simultaneously.

1. Introduction

In today's societies, the uninterrupted supply of electrical energy is essential for the modern human life, where the vast majority of activities are associated directly or indirectly to electricity consumption. Providing electricity energy via smart grid infrastructure is one of the main features of modern cities. Through the smart grid infrastructure, electricity price signal can be propagated to the electricity customers. The signal enables the customers to act, such as active players, who can modify their consumption concerning the electricity price variations. In these conditions, electricity customers' behavior estimation has become an essential concept from different perspectives. Employing the customers' behavior, the independent system operator (ISO) can estimate socio-economic balancing between the electricity production and consumption, effects of tax and subsidy on electricity demand, etc. Informing the customers' behavior, the electricity producers, who operate in a highly competitive environment, can pricing their generation properly. Electricity retailers can provide efficient electrical services and maximize their profit by employing the correct estimation of the customers' consumption behavior. Electricity customers' behavior estimation has become even more critical in recent years due to events like power system deregulation, the rising primary fuel prices, the growing attention of demand-side management programs [1].

One of the main outcomes of the consumers' behavior estimation is modeling price elasticity of electricity demand. Many studies start from single-equation econometric models to estimate electricity demand to analyze consumers' behavior and price elasticity. For example, in [2] log demand model has been used to estimate the price elasticity of electricity demand. In [3] price-elasticity has been studied using log-log demand models for the peak and off-peak hours. Also, the almost ideal demand system (AIDS) model [4-5], ordinary least squares [6], panel data methods [7-8], artificial neural networks [9], and time-series methods [10], [11], [12] and [13] have been utilized to estimate the consumption and the price elasticity data. This paper designs a novel load model based on the concept of the utility function to estimate the customers' behavior and price elasticity.

Although there is a sizeable study on estimation of the price elasticity of electricity demand, many of them have measured country-level aggregated elasticity model rather than the sectoral-level elasticity model [14]. While such estimations could provide comprehensive insights into the customers' behaviors and responses of a particular country, they can't determine the individual behaviors and responses of the customers against the electricity price changes. The individual behavior is often affected by additional benefits of the consumption like productivity, health, convenience, and aesthetics, which might be difficult to quantification [15]. We employ risk aversion coefficients to model the customers' individual risk aversion behavior and estimate the price elasticity of electricity demand.

This paper proposes a novel model to estimate electricity customers' behavior, where the main parameter is the customers' risk aversion coefficients. The coefficients are used to estimate the electricity consumption value from the customers' viewpoints. Where, the concept of the utility function is utilized to determine the customers' welfare function. Then, an economic load model is established based on the welfare function. Using the load model, electricity customer sensitivity against the electricity price and income variations are calculated. The main contributions of this paper are summarized as follows:

- Electricity customers' perspective due to electrical energy consumption is estimated utilizing the customers' risk aversion coefficient.
- The customers' welfare is modeled only as a function of electricity price and the risk aversion coefficients.
- A novel load model is designed based on the risk aversion coefficients.
- A comprehensive elasticity model is developed to estimate the own-price, cross-price, and income elasticity through the risk aversion coefficients.

The remainder of this paper is organized as follows. The next Section reviews the previous studies around the price elasticity of electricity demand and the utility function. Section 3 and 4 describe single and multi-period electricity load model. In these Sections, demand function, welfare function, price elasticity, and income elasticity are formulated considering different utility functions. The numerical results are addressed in Section 5. Finally, the paper is concluded in Section 6.

2. Literature Review

2.1. Price elasticity of electricity demand

In the past years, several authors have published empirical estimations of electricity price elasticity. Ref. [16] has reported different variables that may affect the electricity price elasticity. The variables include sectors of consumption, price of electricity, prices of alternative fuels, climatic variables, heating degree days, cooling degree days, underlying energy demand trend, stock of electric appliances, size of house, size of town, size of region, urbanization rate and population. The studies have estimated the long-run elasticity is higher than the short-run [17]. For example, authors in [18] have studied the short-run and long-run electricity elasticity for the Gulf Cooperation Council (GCC) countries. The short-run and long-run price elasticity for Saudi Arabia, Kuwait, Oman, Bahrain, Qatar and the United Arab Emirates have been estimated at -0.04, -0.08, -0.07, -0.06, -0.18, -0.09 and -1.24, -1.1, -0.82, -3.39, -1.09, -2.43. Also, different electricity sectors may have different elasticities. For example, the own-price elasticity for the Iranian industry, agriculture, and service sectors have been estimated at -0.14, 0, and -0.48, respectively. For this country, the income elasticity has been determined at 0.39, 0.6, and 0.53 for the industry, agriculture, and service sectors [19]. The authors in [10] have illustrated while the industrial consumption is more elastic than the households in the short-run, the elasticity for the households is higher than the industrial customers in the long-run.

The cross-price elasticity of electricity demand has been investigated in some other studies. For example, authors in [4] and [5] have studied the own-price and cross-price elasticity for the peak and off-peak periods. Although the own-price elasticity is always negative, the cross-price elasticity would be positive. Positive cross-price elasticity means some part of the consumption will be shifted to the other period when the electricity price increases.

In addition to the price sensitivity, electricity consumption is sensitive concerning income changes. References [6] and [13] have illustrated the long-run income elasticity for electricity demand is higher than the short-run one. Also, considering electricity demand for the GCC countries, authors in [11] have shown the residential electricity demand is price and income inelastic in the short-run. In contrast, in the long-run, the price and income elasticity are estimated to -0.16 to zero and 0.43 to 0.71. In addition to the price and income elasticity, urbanization elasticity has been considered in [20]. Urbanization elasticity measures how much electricity consumption increases by increasing 1% in the urbanization. The results have illustrated the urbanization elasticity is 1.61 and 3.91 in the short-run and long-run. In Ref. [8], rural and urban areas price elasticity for electricity demand have been compared. The results have illustrated rural customers are more sensitive to price changes. This may happen because they can more easily use other energy sources, for example, wood, for cooking and heating. Also, income elasticity has been reported 0.35 for urban homes and 0.3 for rural ones, which may be explained by the highest dependence on electricity for urban households. Ref. [3] has illustrated that although the own-price elasticity is always negative, the cross-price elasticity can be positive or negative.

Table 1 summarizes the selected literature on the price and income elasticities of energy demand. From the table, the own-price elasticity for electricity demand is inelastic in the short-run (between 0 and -1). . At the same time, it may be inelastic or elastic (lower than -1) in the long-run. Plus, it can be concluded the cross-price elasticity of electricity demand can be negative or positive. The table shows the income elasticity of electricity demand always has a positive value (lower or greater than the unit).

Table 1: Summary of the literature on price and income elasticities for electricity demand.

| Ref. | Country (Sector) | Own-price elasticity | Cross-price elasticity | Income elasticity |
|------|------------------|----------------------|------------------------|-------------------|
| [2]  | Australia (R)    | -0.4165 SR           | -                      | -                 |
| [18] | GCC Countries (A)| -0.04 to -0.18 SR    | -                      | -                 |
|      |                  | -0.82 to -3.39 LR    |                        |                   |
| [14] | OECD Countries (I)| -0.082 t0 -0.186 SR |                        |                   |
|      |                  | -0.2 to -0.669 LR    |                        |                   |
| [10] | Israel (R & I)   | R: -0.124 SR         | -                      | -                 |
|      |                  | R: −0.579 LR         |                        |                   |
|      |                  | I: −0.311 SR         |                        |                   |
|      |                  | I: -0.123 to −0.479 LR |                      |                   |
| [4]  | France (I)       | Off-Peak: -1.87      | peak/off-peak: 0.46    | -                 |
|      |                  | Peak: -1.47          | off-peak/peak: 0.85    |                   |
| [5]  | Switzerland (R)  | Off-peak: −2.30 to −2.57 | 0.34 to 1.57       | -                 |
|      |                  | Peak: −1.25 to −1.41 |                        |                   |
| [10] | GCC Countries (R)| 0 SR                 | -                      | 0 SR              |
|      |                  | 0 to -0.16 LR        |                        | 0.43 to 0.71 LR   |
| [7]  | Japan (R)        | -1.13 LR             | -                      | 0.60 LR           |
| [12] | S.Korea (R)      | -0.27 LR             | -                      | 1.33 LR           |
| [6]  | Pakistan (I)     | -0.06 SR             | -                      | 0.85 SR           |
|      |                  | -0.22 LR             |                        | 2.94 LR           |
| [13] | Turkey (A)       | -0.0123 SR           | -                      | 0.0148 SR         |
|      |                  | -0.9079 LR           |                        | 1.094 LR          |
| [20] | Taiwan (R)       | -0.15                | -                      | 0.23 ST           |
|      |                  |                      |                        | 1.04 LT           |
| [19] | Iran (I, S, a)   | I: -0.14             | I: 0.0 (to Natural Gas)| I: 0.39           |
|      |                  | S: -0.48             | S: 0.38 (to Natural Gas)| S: 0.53          |
|      |                  | a: 0.0               | a: 0.0 (to Natural Gas)| a: 0.6            |
| [8]  | Portugal (R)     | u: -0.673 LR         | u: 0.244 LR            | u: 0.3 LR         |
|      |                  | r: -0.897 LR         | r: 0.375 LR            | r: 0.35 LR        |
| [3]  | India (R)        | −0.452 (Winter)      | −0.27 (Winter)         | 0.64 (Winter)     |
|      |                  | −0.29 (Summer)       | 0.26 (Summer)          | 0.63 (Summer)     |
|      |                  | −0.51 (Monsoon)      | −0.65 (Monsoon)        | 0.60 (Monsoon)    |

SR: Short-run, LR; Long-run, A: Aggregate, I: Industrial, R: Residential, a: Agriculture, S: Service Sector, u: Urban, r: Rural Conaumption.

## 2.2. Utility function

The utility function is a representation to measure individual preferences for goods or services beyond the explicit monetary value of those goods or services. In other words, it is a calculation to estimate customer satisfaction regarding consuming goods or receiving services. Because of its advantages, the utility function is used in some studies to estimate electricity customers' behavior. The main properties of utility functions, which are suitable to express the electricity customers behavior, have been described in [21]. Based on the description, the utility function would be an ascending and concave function, which saturates gradually. The social welfare has been defined in [22] by subtracting electricity costs from the customer utility function. A behavioral real-time pricing method has been established in [23] to encourage electricity customers to participate in the electricity market. In the paper, the concept of the utility function has been utilized to measure the customers' satisfaction due to electrical energy consumption. In Ref. [24], utility function has been utilized to investigate the consumption behavior of the household customers versus the electricity price variations. To maximize wind profit in the power market, Ref. [25] has designed a new incentive-based demand response (DR) program. To specify the incentive rate properly, the quadratic and exponential utility functions have been considered. Novel virtual energy storage has been designed in [26] based on electricity customers' behavior. The objectives of the paper were determining optimal incentive and discount to convince the customers to reduce and increase their consumption, which provide functions similar to storage discharging and charging. A new reliability insurance mechanism has been designed in [27], where different utility functions are employed to estimate electricity outage value from the customers' points of view. Also, a novel risk hedging mechanism based on the utility function has been designed in [28] to protect electricity customers from the market price risks. Table 2 summaries the selected literature on the utility function application in the electricity power system.

**Table 2:** Summary of the literature on utility function aplication in the electricity power system

| Ref. | Utility function Type | Description |
| --- | --- | --- |
| [21] | Quadratic | Determining optimal electricity consumption based on the customers' utility |
| [22] | Quadratic | Formulating social welfare based on the utility function |
| [23] | Quadratic | Designing a behavioral real-time pricing scheme |
| [24] | Cobb-Douglas | Estimating hourly load based on the utility maximization concept |
| [25] | Quadratic, Exponential | Developing incentive-based DR program based on the customers' utility |
| [26] | Quadratic, Exponential | Utilizing DR resources as virtual energy storage considering the customers' behavior |
| [27] | Linear, Quadratic, Exponential | Establishing reliability insurance mechanism considering the customers' and utility company's behavior |
| [28] | Linear, Quadratic, Exponential | Establishing a risk hedging mechanism to protect electricity customers from the electricity market price risks |

## 3. Single-period electricity load model

Some electricity consumptions (e.g., lighting demands) cannot move from one period to another, which can only be turned on or off. In this Section, a new load model for this type of consumptions is developed based on the customers' risk aversion behavior.

### 3.1. Utility function

Utility function represents electricity customer satisfaction, U, concerning the consumption level, D, and his/her risk aversion coefficient, $a$, with the following properties [21]:

Property I: Marginal utility is positive, meaning that the customer's satisfaction increases as the consumption increases, thus:

$$\frac{\partial U(D,a)}{\partial D} \geq 0 \tag{1}$$

Property II: Marginal utility is a non-increasing function, meaning that the marginal satisfaction decreases as the consumption increases, which can be written as:

$$\frac{\partial U^2(D,a)}{\partial D^2} \leq 0 \tag{2}$$

Property III: Zero consumption leads to zero utility, meaning no satisfaction is obtained without the consumption, which can be expressed as:

$$U(0,a) = 0 \tag{3}$$

Property IV: Utility function is non-decreasing with respect to the risk aversion coefficient, meaning that that the customer with a high-risk aversion coefficient obtains more satisfaction due to the consumption, which can be formulated as:

$$\frac{\partial U(D,a)}{\partial a} > 0 \tag{4}$$

Each function with the above properties can be used as a utility function that will concave and saturate by the consumption, as shown in Fig. 1.

The welfare of electricity customers has been formulated in [25] by subtracting electricity cost from the utility function. The calibration coefficient $A$ has been utilized to calibrate the welfare function with the initial consumption and electricity price. The customer welfare versus different electricity prices, $\pi$, is illustrated in Fig.2 and formulated in (5). From the figure, different electricity prices lead to different consumption levels.

$$W(D,\pi,a) = AU(D,a) - \pi D \tag{5}$$

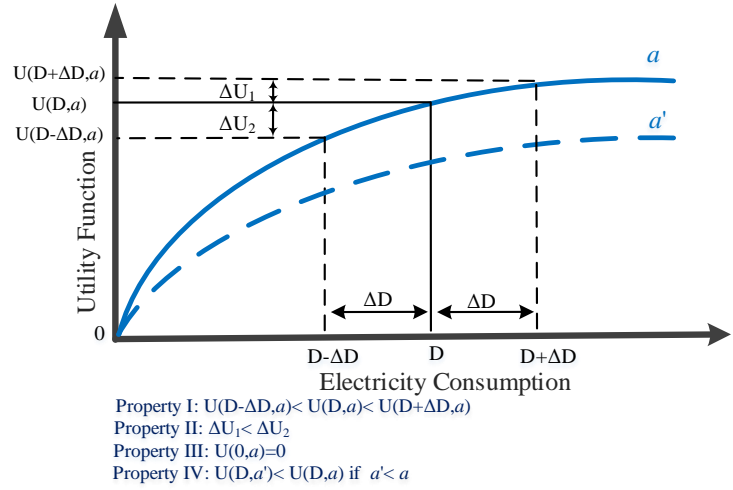

**Fig. 1**: Utility of electrical energy consumption.

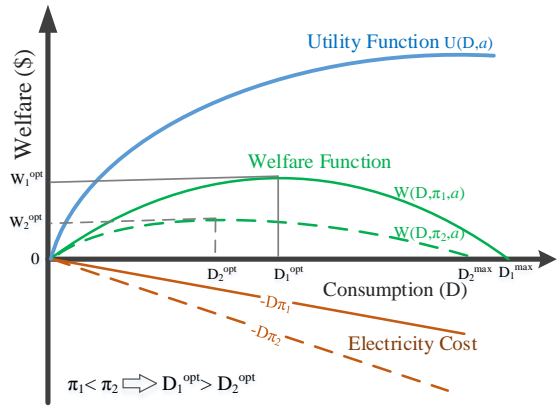

**Fig. 2:** Customer welfare versus different electricity costs.

The customer consumes electric energy to maximize his/her welfare, which can be expressed as:

$$\frac{\partial W(D,\pi,a)}{\partial D} = \frac{\partial (AU(D,a))}{D} - \pi = 0 \tag{6}$$

$$A\frac{\partial U(D,a)}{\partial D} = \pi \tag{7}$$

So, the calibration coefficient would be as (8).

$$A = \frac{\pi_0}{\frac{\partial U(D_0, a)}{\partial D_0}} \tag{8}$$

where $\pi_0$ and $D_0$ are the initial electricity price and initial consumption level. The utility function can measure the customers' risk aversion behavior in which the higher the curvature of U indicates the higher the risk aversion behavior. Since, several functions can be used as a utility function, a measure that stays constant concerning these functions is needed. One such measure is the Arrow–Pratt measure of absolute risk aversion (ARA), defined as [29]:

$$ARA(D) = -\frac{\partial^2 U(D,a)/\partial D^2}{\partial U(D,a)/\partial D} \tag{9}$$

From table 2, the linear, quadratic, and exponential forms of utility function are the most widely used functions to describe electricity customers' behavior. Authors in [27] have illustrated the linear utility function leads to infinite elasticity, which is unreasonable for electricity consumption. Hence, the authors have concluded the linear utility function is unsuitable for modeling electricity customers' behavior. Therefore, this paper considers the exponential and the quadratic utility functions to model electricity customers' behavior and estimate the price elasticity of electricity demand.

### 3.1.1. Exponential utility function

The exponential utility function has been formulated as follows in [25].

$$U(D,a) = 1 - e^{-aD}, \quad a > 0 \tag{10}$$

For the exponential utility function, the welfare function, the calibration coefficient, and the Arrow–Pratt measure would be as follows:

$$W(D,\pi,a) = A(1 - e^{-aD}) - \pi D \tag{11}$$

$$A = \frac{\pi_0}{a} e^{aD_0} \tag{12}$$

$$ARA(D) = a \tag{13}$$

It is obvious the electricity consumption cannot be negative so, the electricity consumption would be as in (14) by substituting (10) in (6).

$$\frac{\partial W(D,a)}{\partial D} = \frac{\partial [A(1-e^{-aD}) - \pi D]}{\partial D} = 0 \tag{14}$$

$$Aae^{-aD} - \pi = 0 \tag{15}$$

$$D = \begin{cases} D_0 - \frac{1}{a}\ln(\frac{\pi}{\pi_0}), & \pi < \pi_0 e^{aD_0} \\ 0, & \pi \geq \pi_0 e^{aD_0} \end{cases} \tag{16}$$

From (16), as the risk aversion coefficient increases, the consumption decreases further by the price increment.

The welfare function also can be expressed only as a function of the electricity price and the risk aversion coefficient by substituting (16) in (11), which is represented in (17) and illustrated in Fig. 3.

$$W_\pi(\pi,a) = \begin{cases} A + \frac{\pi}{a}[\ln(\frac{\pi}{aA}) - 1], & \pi < \pi_0 e^{aD_0} \\ 0, & \pi \geq \pi_0 e^{aD_0} \end{cases} \tag{17}$$

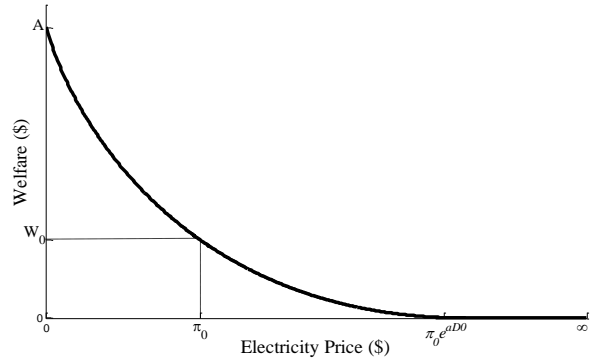

**Fig. 3**: Customer welfare versus different electricity prices for the exponential utility function.

### 3.1.2. Quadratic utility function

The quadratic utility function has been formulated in [25] as follows:

$$U(D,a) = -(1-aD)^2, \quad D \leq \frac{1}{a} \tag{18}$$

For this form of the utility function, the welfare function, the calibration coefficient, the consumption function, and the Arrow–Pratt measure would be as follows:

$$W(D,\pi,a) = -A(1-aD)^2 - \pi D \tag{19}$$

$$A = \frac{\pi_0}{2a(1-aD_0)} \tag{20}$$

$$D = \begin{cases} \dfrac{1}{a} - \dfrac{\pi}{\pi_0}\dfrac{(1-aD_0)}{a}, & \pi < \dfrac{\pi_o}{1-aD_0} \\ 0, & \pi \geq \dfrac{\pi_o}{1-aD_0} \end{cases} \tag{21}$$

$$ARA(D) = \frac{a}{(1-aD)} \tag{22}$$

Also, equation (23) and Fig. 4 illustrate the customer welfare only as a function of electricity price for the quadratic utility function. It should be mentioned the welfare and utility are relative concepts to compare different consumption values. Hence, they can be assigned to the negative values.

$$W_\pi(\pi, a) = \begin{cases} \dfrac{\pi^2}{4a^2 A} - \dfrac{\pi}{a}, & \pi < \dfrac{\pi_o}{1-aD_0} \\ -A, & \pi \geq \dfrac{\pi_o}{1-aD_0} \end{cases} \tag{23}$$

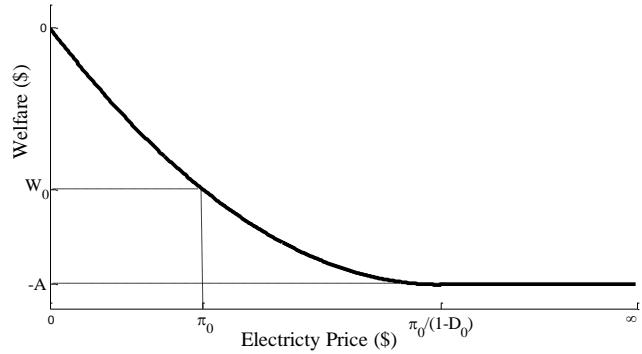

**Fig. 4**: Customer welfare against different electricity prices for the quadratic utility function.

### 3.2. Price-elasticity for single-period demands

Price-elasticity of electricity demand is a normalized measurement of the proportional change of the customer's consumption in response to a change in the electricity price. For single-period demands, only own-price elasticity can be defined in which the demand of *I*th period responds only to the *I*th price variations. The own-price elasticity would be as in (24).

$$E_I = \frac{\pi_I}{D_I} \cdot \frac{\partial D_I}{\partial \pi_I}\bigg|_{D_I = D_I^0, \pi_I = \pi_I^0} \tag{24}$$

In the following, own-price elasticity of the single-period electricity demand will be determined considering different utility functions.

### 3.2.1. Exponential utility function

Considering (16) as the electricity consumption function for the exponential utility function, the derivation of consumption concerning the electricity price would be as (25). Considering (24) and (25), the own-price elasticity for the exponential utility function would be as (26). Due to the exponential utility function definition, $a \geq 0$, the own-price elasticity always will be negative.

$$\frac{\partial D_I}{\partial \pi_I} = -\frac{1}{a_I \pi_I} \tag{25}$$

$$E_I = -\frac{1}{a_I D_I^0} \tag{26}$$

Based on (26), the risk aversion coefficient plays an essential role in the customer's behavior against the price variations. Table 3 illustrates different own-price elasticities concerning the value of $a_I D_I^0$, which are described in the following:

a) Perfect inelastic demand: Infinite risk aversion coefficient leads to perfect inelastic demand. This type of demand remains constant for any value of the price. It is practically impossible to find a load that has a perfectly inelastic demand.

b) Inelastic demand: Demand is inelastic when a change in the price causes a relatively small effect on the demand. To this end, the term $a_I D_I^0$ has to be greater than the unit.

c) Unit elastic demand: Demand is unit elastic when the percentage change in demand is equal to the percentage change in the price. This happened when $a_I D_I^0$ equals to the unit.

d) Elastic demand: Demand is elastic when a change in the price produces a relatively enormous effect on the demand. Demand is elastic when $a_I D_I^0$ is lower than the unit.

e) Perfect elastic demand: In this case, the customer is risk-neutral due to electrical energy consumption ($a=0$). Where, a small rise (fall) in electricity price leads to zero (infinite) demand. This is also a theoretical concept because it requires the consumer consumes electricity energy all available at some price, but none at any other price.

**Table 3:** Own-price elasticity as a function of $a_I D_I^0$ for exponential utility function.

| | | |
|---|---|---|
| Perfect inelastic demand | $a_I D_I^0 = \infty$ | $E_I = 0$ |
| Inelastic demand | $a_I D_I^0 > 1$ | $0 < E_I < -1$ |
| Unit elastic demand | $a_I D_I^0 = 1$ | $E_I = -1$ |
| Elastic demand | $a_I D_I^0 < 1$ | $-\infty < E_I < -1$ |
| Perfect elastic demand | $a_I D_I^0 = 0$ | $E_I = -\infty$ |

### 3.2.2. Quadratic utility function

Considering (21), the own-price elasticity for the quadratic utility function would be as follows:

$$E_I = \frac{-(1 - a_I D_I^0)}{a_I D_I^0} \tag{27}$$

### 3.2.3. General form of own-price elasticity

Considering the Arrow–Pratt measure, the general form of the own-price elasticity would be as follows [27]:

$$E_I = -\frac{1}{D_I^0 ARA(D_I^0)} \tag{28}$$

Based on the above descriptions, the risk aversion coefficients and price elasticities are related to each other. The risk aversion coefficient can be expressed based on the elasticity data, as shown in (29) and (30) for the exponential and quadratic utility functions.

$$a_I = \frac{-1}{D_I^0 E_I} \tag{29}$$

$$a_I = \frac{-1}{D_I^0 (E_I - 1)} \tag{30}$$

## 4. Multi-period electricity load model

In addition to own price sensitivity, some demands are sensitive to price changes in the other periods. For example, to reduce the electricity bill, consumption (e.g., washing machine) can be transferred from the peak to the off-peak period. Hence, the consumption will be related to the different periods' price. The customer welfare for the multi-period demand will be a function of different time periods' utility and electricity cost, which can be expressed as follows:

$$W_T = A_1 U(D_1, a_1) + ... + A_T U(D_T, a_T) - [\pi_1 D_1 + ... + \pi_T D_T] \tag{31}$$

The calibration coefficients can be determined maximizing the welfare function concerning the initial demands as bellows:

$$\left. \frac{\partial W_T}{\partial D_I} \right|_{D_I = D_I^0} = 0, \quad \forall I \tag{32}$$

$$A_I = \left. \frac{\pi_I}{\frac{\partial U(D_I, a_I)}{\partial D_I}} \right|_{D_I = D_I^0, \pi_I = \pi_I^0}, \quad \forall I \tag{33}$$

where the calibration coefficient at each period is only a function of the initial demand and electricity price of that period.

### 4.1. Price-elasticity for multi-period demands

For multi-period demands, the consumption has multi-period sensitivity, which is denoted by cross-price elasticity of demand. The cross-price elasticity of $I$th period demand concerning the $J$th period price can be formulated as in (34).

$$E_{I,K} = \left. \frac{\pi_K}{D_I} \cdot \frac{\partial D_I}{\partial \pi_K} \right|_{D_I = D_I^0, \pi_K = \pi_K^0} \tag{34}$$

#### 4.1.1. Exponential utility function

Solving optimization (35), which is a Lagrangian optimization problem, leads to the customer consumption formula at each time period. The constrain of this optimization is illustrated in (36), which means the sum of the electricity costs for the different periods has to be equal to the customer budget ($B$).

$$Max: W_T \tag{35}$$

S.T:

$$\pi_1 D_1 + ... + \pi_T D_T = B \tag{36}$$

Solving the above optimization for the exponential utility function, we have:

$$\begin{aligned} A_I a_I e^{-a_I D_I} &= \lambda \pi_I \\ &\vdots \\ A_T a_T e^{-a_T D_T} &= \lambda \pi_T \end{aligned} \tag{37}$$

where $\lambda$ is the Lagrange multiplier that can be interpreted as the incremental satisfaction of electrical energy consumption. From the Lagrangian optimization, the Lagrange multiplier will be constant in the different periods, so we have:

$$\frac{A_I a_I e^{-a_I D_I}}{\pi_I} = ... = \frac{A_T a_T e^{-a_T D_T}}{\pi_T} = \lambda \tag{38}$$

$$\begin{aligned} D_2 &= \frac{a_I}{a_2} D_I + \frac{1}{a_2} \ln(\frac{\pi_I}{\pi_2} \frac{A_2}{A_I} \frac{a_2}{a_I}) \\ &\vdots \\ D_T &= \frac{a_I}{a_T} D_I + \frac{1}{a_T} \ln(\frac{\pi_I}{\pi_T} \frac{A_T}{A_I} \frac{a_T}{a_I}) \end{aligned} \tag{39}$$

As shown in (39), different time periods' consumption are related to each other. Substituting (39) in (36), the demand for the *I*th period is determined as in (40).

$$D_I = \frac{B - \sum_{J=1}^{T} \frac{\pi_J}{a_J} \ln(\frac{\pi_I}{\pi_J} \frac{a_J}{a_I} \frac{A_J}{A_I})}{\sum_{J=1}^{T} \frac{a_I}{a_J} \pi_J} \tag{40}$$

Considering the first derivative of (40), the own-price elasticity would be as follows:

$$E_{I,I} = -\frac{1}{a_I D_I^0} + \frac{1 - a_I D_I^0}{a_I D_I^0 \sum_{J=1}^{T} \frac{\pi_J^0}{\pi_I^0} \frac{a_I}{a_J}} \tag{41}$$

Based on (41), the own-price elasticity for the multi-period demand may be larger, equal, or smaller than the single-period model. The cross-price elasticity for exponential utility function would be as follows:

$$E_{I,K} = \frac{(1-a_I D_I^0) - \ln(\frac{\pi_I^0}{\pi_K^0} \frac{A_K}{A_I} \frac{a_K}{a_I})}{a_I D_I^0 \sum_{J=1}^{T} \frac{\pi_J^o}{\pi_K^o} \frac{a_K}{a_J}} \tag{42}$$

Substituting $A_I$ and $A_K$ from (33) in (42), we have:

$$E_{I,K} = \frac{1}{a_I D_I^0} \frac{(1-a_K D_K^0)}{\sum_{J=1}^{T} \frac{\pi_J^0}{\pi_K^0} \frac{a_K}{a_J}} \tag{43}$$

As shown in Table 4, being consumption substitutes or complement between different periods depends on $a_K D_K^0$, which are discussed in the following.

a) Substitute demand: Substitute demand is consumption with a positive cross-price elasticity, meaning the demand increases when the price of another period increases.

b) Independent demand: Independent demand is consumption with a zero cross-price elasticity, meaning the demand does not change when the price of another period changes.

c) Complementary demand: Complementary demand is a consumption with a negative cross-price elasticity, meaning the demand increases when the price of another time period decreases.

**Table 4:** Cross-price elasticity for exponential utility function as a function of $a_K D_K^0$

| | | |
|---|---|---|
| Substitute demands | $a_K D_K^0 < 1$ | $E_{I,J} > 0$ |
| Independent demands | $a_K D_K^0 = 1$ | $E_{I,J} = 0$ |
| Complementary demands | $a_K D_K^0 > 1$ | $E_{I,J} < 0$ |

### 4.1.2. Quadratic utility function

Similar to the exponential utility function, electricity consumption, own-price elasticity, and cross-price elasticity for the quadratic utility function would be as follows:

$$D_I = \frac{B - \sum_{J=1}^{T} [\frac{\pi_J}{a_J} - \frac{A_I}{A_J} (\frac{\pi_J}{a_J})^2 \frac{a_I}{\pi_I}]}{\sum_{J=1}^{T} \frac{\pi_J^2}{\pi_I} \frac{A_I}{A_J} (\frac{a_I}{a_J})^2} \tag{44}$$

$$E_{I,I} = -\frac{(1-a_I D_I^0)}{a_I D_I^0} + \frac{(1-2a_I D_I^0)}{a_I D_I^0 \sum_{J=1}^{T} (\frac{\pi_J^0}{\pi_I^0})^2 \frac{A_I}{A_J} (\frac{a_I}{a_J})^2} \tag{45}$$

$$E_{I,K} = \frac{1-a_I D_I^0}{a_I D_I^0} \frac{2}{\sum_{J=1}^{T} (\frac{\pi_J^0}{\pi_k^0})^2 \frac{A_K}{A_J} (\frac{a_k}{a_J})^2} - \frac{\pi_k^0}{a_K D_I^0 \sum_{J=1}^{T} \frac{\pi_J^{0^2}}{\pi_I^0} \frac{A_I}{A_J} (\frac{a_I}{a_J})^2} \tag{46}$$

### 4.2. Income-elasticity of demand

Income elasticity is an economic term that explains the connection between demand and the consumer's income. In other words, if a customer's income goes up or down, his/her income elasticity represents how he/she will use the electrical energy. In the following, the income-elasticity of electricity demand will be formulated for the different utility functions.

#### 4.2.1. Exponential utility function

The income elasticity is the ratio of the percentage change in demand to the percentage change in income. The first derivation of the consumption concerning $B$ and income-elasticity of demand for the exponential utility function are illustrated in (47) and (48).

$$\frac{\partial D_I}{\partial B} = \frac{1}{\sum_{J=1}^{T} \frac{a_I}{a_J} \pi_J} \tag{47}$$

$$E_{I,B} = \frac{B}{a_I D_I^0 \sum_{J=1}^{T} \frac{\pi_J^0}{a_J}} \tag{48}$$

Table 5 compares different income elasticities which are described in the following:

a) Inferior demand: An inferior demand has negative income elasticity. This means the demand will decrease as the consumer's income increases. Inferior demands are called inferior because they usually have superior alternatives. For example, if a consumer's income increases, he/she might start using LED lights instead of incandescent lights, which needs less electricity consumption.

b) Normal demand: A normal demand has an income-elasticity between zero and one. This means the demand will increase as the consumer's income increases. The percentage of change in the demand is less in proportion to the percentage of change in consumers' income. Normal demand includes basic electricity demand. For example, living room lighting is a normal lighting demand.

c) Luxury demand: Luxury demand usually has income-elasticity greater than one, which means it is income elastic. The percentage of change in the demand is more in proportion to the percentage of change in consumers' income. For instance, if a consumer's income increases, he/she might invest in luxury lighting.

Table 5: Different types of income elasticity.

| | |
|---|---|
| $E_{I,B} < 0$ | Inferior demand |
| $0 < E_{I,B} < 1$ | Normal demand |
| $E_{I,B} > 1$ | Luxury demand |

### 4.2.2. Quadratic utility function

Similar to the exponential utility function, income elasticity for the quadratic utility function would be as follows:

$$E_{i,B} = \frac{B}{D_I^0 \sum_{J=1}^{T} \frac{\pi_J^{02}}{\pi_I^0} \frac{A_I}{A_J} (\frac{a_I}{a_J})^2} \tag{49}$$

### 4.3. Relation between price elasticity and income elasticity

Comparing the income elasticity with the price elasticity, the following equilibrium can be extracted.

$$\sum_{j=1}^{T} E_{i,j} = -E_{i,B} \tag{50}$$

The equilibrium implies income elasticity is related to the price elasticity. Indeed, the consumption decrement due to 1 percent electricity price increment in all periods, would be the same with 1 percent income decrement.

Tables 6 and 7 summarized the customers' behavior equations for the single and multi-period load models. The main parameter of these questions is the customers' risk aversion coefficient.

Table 6: Summary of the customers' behavior equations for the single-period load model

| Exponential Utility Function | Quadratic Utility Function |
|---|---|
| $U(D,a) = 1 - e^{-aD}, \quad a > 0$ | $U(D,a) = -(1-aD)^2, \quad D \leq \dfrac{1}{a}$ |
| $D = \begin{cases} D_0 - \dfrac{1}{a}\ln(\dfrac{\pi}{\pi_0}), & \pi < \pi_0 e^{aD_0} \\ 0, & \pi \geq \pi_0 e^{aD_0} \end{cases}$ | $D = \begin{cases} \dfrac{1}{a} - \dfrac{\pi}{\pi_0}\dfrac{(1-aD_0)}{a}, & \pi < \dfrac{\pi_o}{1-aD_0} \\ 0, & \pi \geq \dfrac{\pi_o}{1-aD_0} \end{cases}$ |
| $W(D,\pi,a) = A(1-e^{-aD}) - \pi D$ | $W(D,\pi,a) = -A(1-aD)^2 - \pi D$ |
| $W_\pi(\pi,a) = \begin{cases} A + \dfrac{\pi}{a}[\ln(\dfrac{\pi}{aA})-1], & \pi < \pi_0 e^{aD_0} \\ 0, & \pi \geq \pi_0 e^{aD_0} \end{cases}$ | $W_\pi(\pi,a) = \begin{cases} \dfrac{\pi^2}{4a^2 A} - \dfrac{\pi}{a}, & \pi < \dfrac{\pi_o}{1-aD_0} \\ -A, & \pi \geq \dfrac{\pi_o}{1-aD_0} \end{cases}$ |
| $E_I = -\dfrac{1}{a_I D_I^0}$ | $E_I = \dfrac{-(1-a_I D_I^0)}{a_I D_I^0}$ |

U: Utility function, D: Load model, W: Welfare function, $W_\pi$: Welfare function with respect only the electricity price and risk aversion coefficient, $E_I$: Price elasticity of electricity demand.

Table 7: Summary of the customers' behavior equations for the multi-period load model

| Exponential Utility Function | Quadratic Utility Function |
|---|---|
| $D_I = \dfrac{B - \sum_{J=1}^{T}\dfrac{\pi_J}{a_J}\ln(\dfrac{\pi_I}{\pi_J}\dfrac{a_J}{a_I}\dfrac{A_J}{A_I})}{\sum_{J=1}^{T}\dfrac{a_I}{a_J}\pi_J}$ | $D_I = \dfrac{B - \sum_{J=1}^{T}[\dfrac{\pi_J}{a_J} - \dfrac{A_I}{A_J}(\dfrac{\pi_J}{a_J})^2 \dfrac{a_I}{\pi_I}]}{\sum_{J=1}^{T}\dfrac{\pi_J^2}{\pi_I}\dfrac{A_I}{A_J}(\dfrac{a_I}{a_J})^2}$ |
| $E_{I,I} = -\dfrac{1}{a_I D_I^0} + \dfrac{1-a_I D_I^0}{a_I D_I^0 \sum_{J=1}^{T}\dfrac{\pi_J^0}{\pi_I^0}\dfrac{a_I}{a_J}}$ | $E_{I,I} = -\dfrac{(1-a_I D_I^0)}{a_I D_I^0} + \dfrac{(1-2a_I D_I^0)}{a_I D_I^0 \sum_{J=1}^{T}(\dfrac{\pi_J^0}{\pi_I^0})^2 \dfrac{A_I}{A_J}(\dfrac{a_I}{a_J})^2}$ |
| $E_{I,K} = \dfrac{1}{a_I D_I^0}\dfrac{(1-a_K D_K^0)}{\sum_{J=1}^{T}\dfrac{\pi_J^0}{\pi_K^0}\dfrac{a_K}{a_J}}$ | $E_{I,K} = \dfrac{1-a_I D_I^0}{a_I D_I^0}\dfrac{2}{\sum_{J=1}^{T}(\dfrac{\pi_J^0}{\pi_k^0})^2 \dfrac{A_K}{A_J}(\dfrac{a_k}{a_J})^2} - \dfrac{\pi_k^0}{a_K D_I^0 \sum_{J=1}^{T}\dfrac{\pi_J^{02}}{\pi_I^0}\dfrac{A_I}{A_J}(\dfrac{a_I}{a_J})^2}$ |
| $E_{I,B} = \dfrac{B}{a_I D_I^0 \sum_{J=1}^{T}\dfrac{\pi_J^0}{a_J}}$ | $E_{i,B} = \dfrac{B}{D_I^0 \sum_{J=1}^{T}\dfrac{\pi_J^{02}}{\pi_I^0}\dfrac{A_I}{A_J}(\dfrac{a_I}{a_J})^2}$ |

D: Load model, $E_{I,I}$: Own-price elasticity, $E_{I,K}$: Cross-price elasticity, $E_{I,B}$: Income elasticity.

## 5. Numerical results

To evaluate the performance of the proposed load model, the Iranian power system load on 17 December 2019, Fig. 5, has been selected for simulation studies [30]. The load curve is divided into three different periods, namely the off-peak period (0:00–8:00), mid-peak period (08:00–17:00), and peak period (17:00–23:00). Table 8 illustrates different periods and corresponding electricity prices for the selected power system [31].

In this Section only the exponential utility function is selected to study the electricity customers' behavior. It should be mentioned the similar results would be achieved utilizing any other form of the utility function. The customers' risk aversion coefficients play an essential role in the proposed load model. Social studies are needed to estimate these coefficients. Utilizing (29), the risk aversion coefficients also can be determined by employing the elasticity data. Table 9 illustrates elasticity data for different Iranian electricity sectors which are taken from [32].

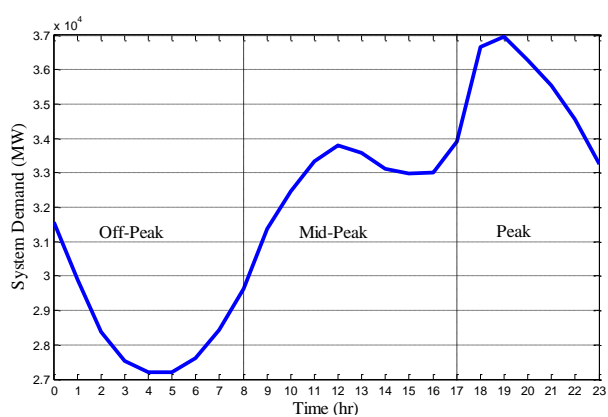

**Fig. 5**. Iranian power grid load curve on 17 December 2019.

**Table 8**: Electricity price data for different periods [31]

|  | Off-peak | Mid-peak | peak |
|---|---|---|---|
| Time | 00:00 to 8:00 | 8:00 to 17:00 | 17:00 to 23:00 |
| Electricity price | 260 Rial/kW | 520 Rial/kW | 1040 Rial/kW |

**Table 9**: Elasticity data for different electricity sectors [32]

| Sector | Off-peak | Mid-peak | peak | Portion of total demand |
|---|---|---|---|---|
| Residential | 1.21 | 0.64 | 1.01 | 34% |
| Industrial | 1.72 | 4.5 | 0.53 | 31% |
| Agricultural | 0.87 | 0.46 | 0.56 | 17% |
| Public | 1.57 | 2.87 | 1.05 | 9% |
| Commercial | 1.1 | 1.99 | 1.6 | 9% |

Utilizing (29), risk aversion coefficients for different customers can be determined, as shown in Table 10. Critical data are embedded in these coefficients. The customers' consumption behavior, achieved utility, and welfare are some of these data. Different consumptions lead to different satisfactions and utility. Figures 6-8 illustrate the customers' utility due to different consumption levels. From the figures, the customers' utility increases as the consumption increases. In the off-peak, the achieved utility is close to each other for different sectors. While in the mid-peak, the agricultural and residential sectors' utility saturate rapidly, the industrial, public, and commercial sectors' utility increases almost linearly. Although the industrial customers have the lowest utility at the off-peak and mid-peak periods, the highest utility would be achieved by this sector in the peak period.

**Table 10**: Risk aversion coefficient for different electricity sectors.

| Sector | Off-peak | Mid-peak | peak |
|---|---|---|---|
| Residential | 0.82 | 1.56 | 0.99 |
| Industrial | 0.58 | 0.22 | 1.89 |
| Agricultural | 1.15 | 2.17 | 1.78 |
| Public | 0.64 | 0.35 | 0.95 |
| Commercial | 0.91 | 0.5 | 0.62 |

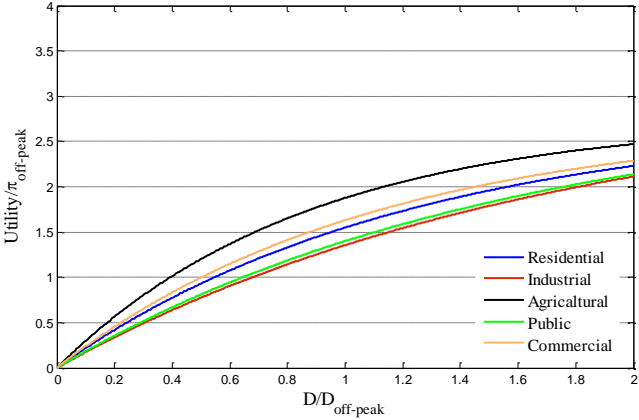

**Fig. 6**: Achieved utility versus different consumption levels in the off-peak period.

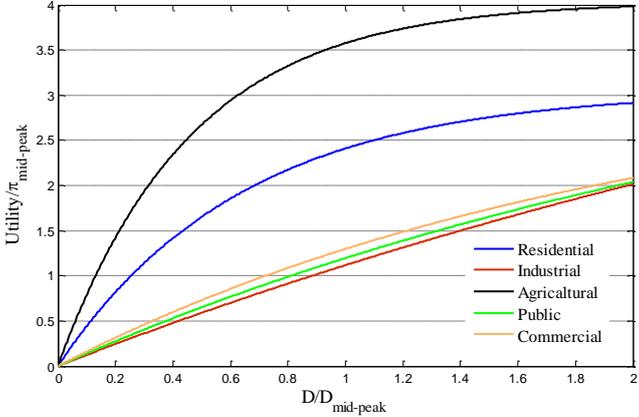

**Fig. 7**: Achieved utility versus different consumption levels in the mid-peak period.

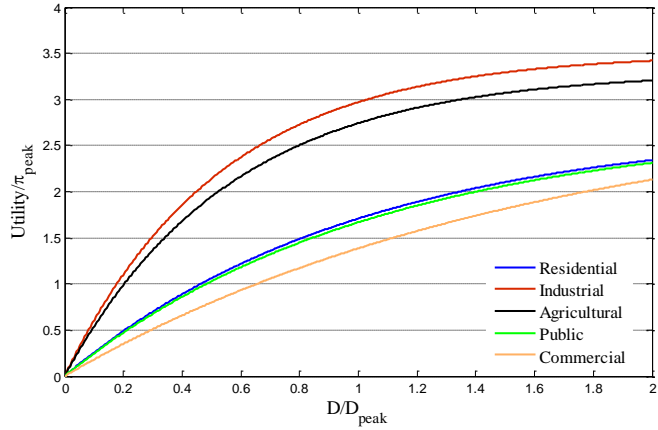

**Fig. 8**: Achieved utility versus different consumption levels in the peak period.

To maximize his/her welfare, electricity customer modifies consumption concerning the electricity price. Figures 9-11 illustrate the customers' consumption versus different electricity prices in which, the consumption decreases as the price increases. In the off-peak, the consumption reduction is almost similar for different sectors. While in the mid-peak, the agricultural and industrial consumptions have the lowest and highest consumption variation. In this period, the industrial, public, and commercial customers consume electricity energy till the price be less than 1.25, 1.4, and 1.65 times bigger than the initial mid-peak price, respectively. In the peak period, consumption of the residential/public and industrial/agriculture sectors are so close to each other.

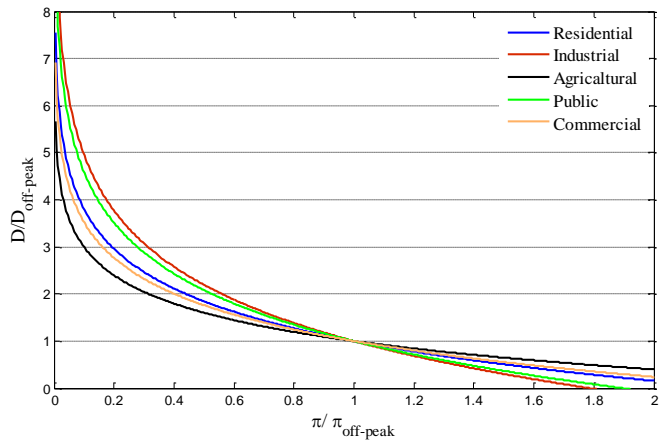

**Fig. 9**: Electricity consumption versus different electricity prices in the off-peak period.

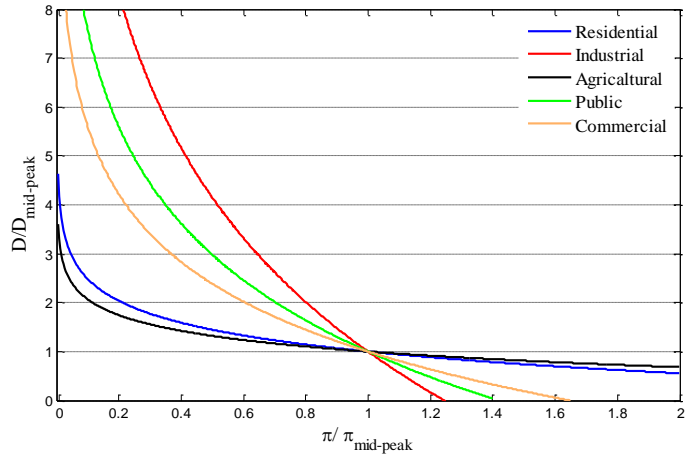

**Fig. 10**: Electricity consumption versus different electricity prices in the mid-peak period.

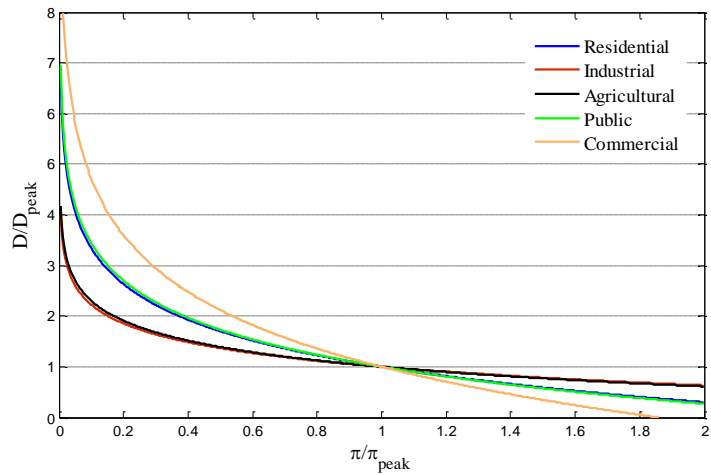

**Fig. 11**: Electricity consumption versus different electricity prices in the peak period.

Figures 12-14 show the customers' welfare concerning different electricity prices. From the figures, welfare decreases as the electricity price increases. It can be seen the agricultural sector obtains the highest welfare from electrical energy consumption. Although the industrial sector achieves the lowest welfare in the off-peak, it earns the highest welfare in the peak period.

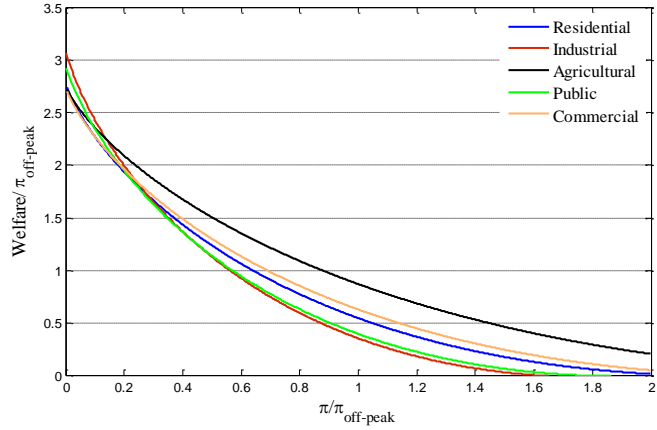

**Fig. 12**: Achieved welfare versus different electricity prices in the off-peak period.

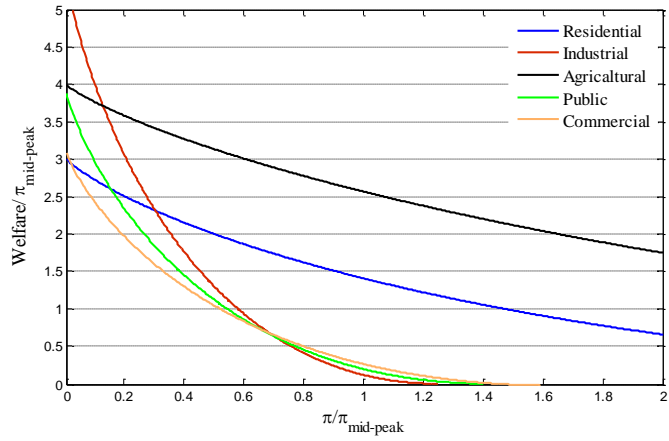

**Fig. 13**: Achieved welfare versus different electricity prices in the mid-peak period.

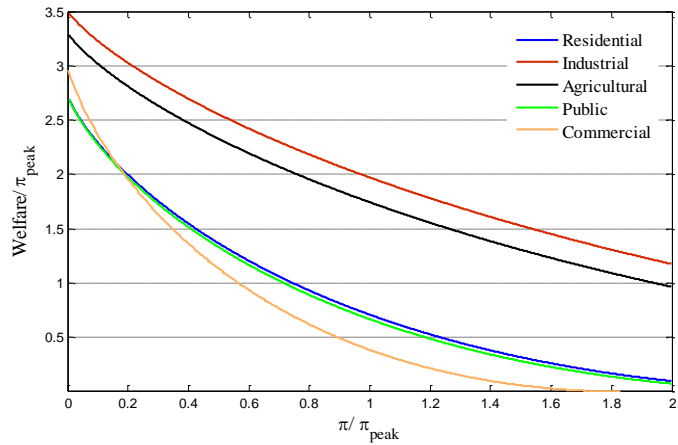

**Fig. 14**: Achieved welfare versus different electricity prices in the peak period.

In the previous Section a comprehensive load model is developed based on the customers' risk aversion coefficients. In addition to own-price elasticity, the model is also capable of estimating cross-price and income elasticity for electricity demand. The assumption that the risk aversion coefficients for the multi-period load model are the same as Table 10, Table 11 illustrates the price and income elasticity matrices for different sectors. The table includes important data about the cross-price elasticity of electricity demand. From the table, while the own-price elasticity is always negative, the cross-price elasticity may be negative or positive, which means different periods' demands can be complements or substitutes. The consumption will be complement (substitute) if the risk aversion coefficient is greater (lower) than the unit. For the unit risk aversion coefficient, the cross-price elasticity will be zero, which means the consumption is independent of the other period's demand. Like the price elasticity, the income elasticity is related to the electricity price and the customers' risk aversion coefficients in different periods. As an example, although the consumption for the industrial sector is highly sensitive to the income variations in the mid-peak, the sensitivity would be small in the peak period.

**Table 11**: Elasticity matrices considering multi-period load model.

| Sector | $E$ | $E_B$ |
|---|---|---|
| Residential | $\begin{array}{c} \\ off-peak \\ mid-peak \\ peak \end{array} \begin{array}{c} off-peak \quad mid-peak \quad peak \\ \begin{bmatrix} -1.18 & -0.13 & 0.007 \\ 0.021 & -0.71 & 0.004 \\ 0.034 & -0.11 & -1.004 \end{bmatrix} \end{array}$ | $\begin{array}{c} off-peak \quad mid-peak \quad peak \\ \begin{bmatrix} 1.3 & 0.68 & 1.08 \end{bmatrix} \end{array}$ |
| Industrial | $\begin{array}{c} \\ off-peak \\ mid-peak \\ peak \end{array} \begin{array}{c} off-peak \quad mid-peak \quad peak \\ \begin{bmatrix} -1.63 & 0.945 & -0.251 \\ 0.25 & -2.050 & -0.664 \\ 0.029 & 0.290 & -0.606 \end{bmatrix} \end{array}$ | $\begin{array}{c} off-peak \quad mid-peak \quad peak \\ \begin{bmatrix} 0.933 & 2.4 & 0.286 \end{bmatrix} \end{array}$ |
| Agricultural | $\begin{array}{c} \\ off-peak \\ mid-peak \\ peak \end{array} \begin{array}{c} off-peak \quad mid-peak \quad peak \\ \begin{bmatrix} -0.898 & -0.23 & -0.38 \\ -0.015 & -0.58 & -0.20 \\ -0.018 & -0.15 & -0.80 \end{bmatrix} \end{array}$ | $\begin{array}{c} off-peak \quad mid-peak \quad peak \\ \begin{bmatrix} 1.50 & 0.80 & 0.97 \end{bmatrix} \end{array}$ |
| Public | $\begin{array}{c} \\ off-peak \\ mid-peak \\ peak \end{array} \begin{array}{c} off-peak \quad mid-peak \quad peak \\ \begin{bmatrix} -1.48 & 0.505 & 0.029 \\ 0.140 & -1.93 & 0.052 \\ 0.051 & 0.340 & -1.03 \end{bmatrix} \end{array}$ | $\begin{array}{c} off-peak \quad mid-peak \quad peak \\ \begin{bmatrix} 0.952 & 1.74 & 0.641 \end{bmatrix} \end{array}$ |
| Commercial | $\begin{array}{c} \\ off-peak \\ mid-peak \\ peak \end{array} \begin{array}{c} off-peak \quad mid-peak \quad peak \\ \begin{bmatrix} -1.09 & 0.195 & 0.21 \\ 0.018 & -1.64 & 0.38 \\ 0.013 & 0.27 & -1.24 \end{bmatrix} \end{array}$ | $\begin{array}{c} off-peak \quad mid-peak \quad peak \\ \begin{bmatrix} 0.684 & 1.24 & 0.957 \end{bmatrix} \end{array}$ |

To investigate the customers' consumption behavior due to electricity price and income variations, three scenarios are considered in this Section. The scenarios are:

Scenario I: Only the impact of electricity price variations on consumption is considered.

Scenario II: Only the impact of income variations on consumption is considered.

Scenario III: The simultaneous impact of electricity price and income variations on consumption is considered.

**Scenario I**

In this scenario, the Iranian power system demand variations versus electricity price changes will be studied. Figures 15-17 illustrate the demand, when the price increases 5% in the off-peak, mid-peak, and peak periods, respectively. The price increment in each period leads to the consumption decrement in that period. In addition to the load decrement, the price increment may affect the other periods' consumption slightly. While the other periods' consumption will increases when the price increases in the off-peak and mid-peak, it decreases by the price increment in the peak period.

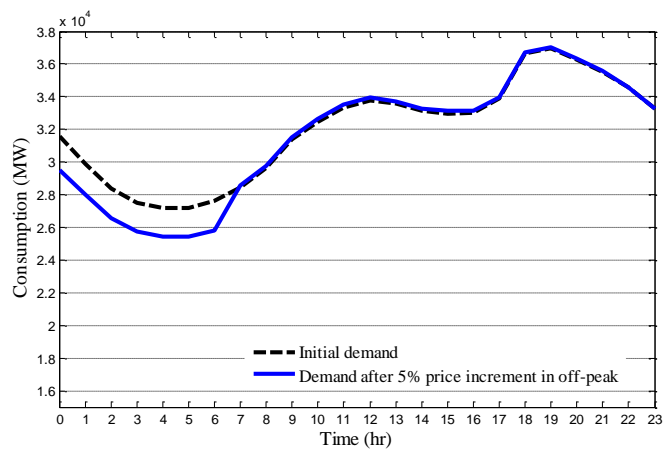

**Fig. 15**: Electricity demand variations versus 5% of electricity price increment in the off-peak period.

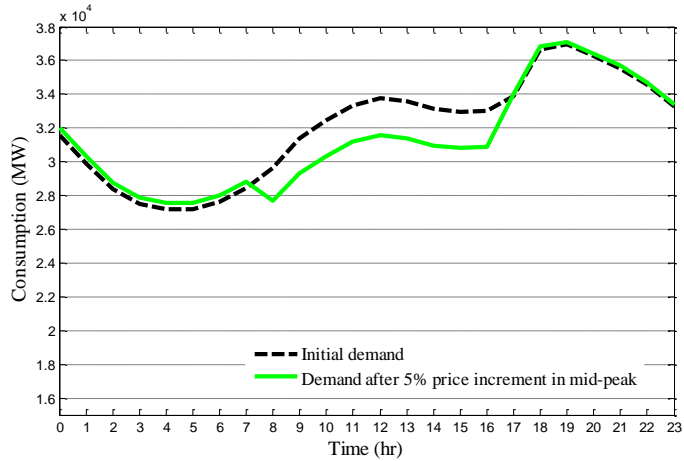

**Fig. 16**: Electricity demand variations versus 5% of electricity price increment in the mid-peak period.

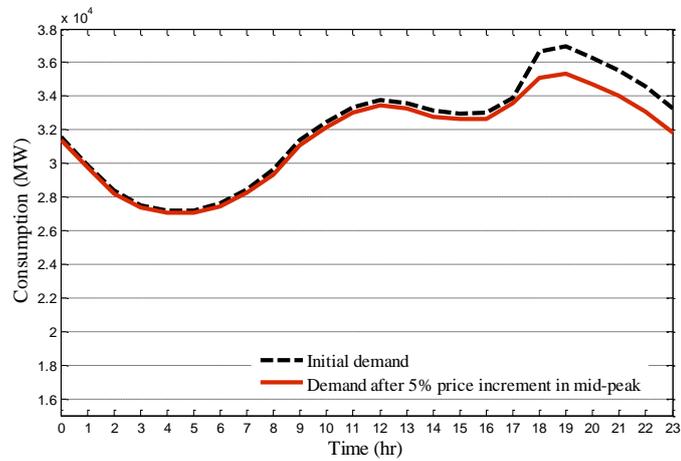

**Fig. 17**: Electricity demand variations versus 5% of electricity price increment in the peak period.

**Scenario II**

This scenario is devoted to studying the income variations effects on the Iranian electricity sectors. To this end, figure 18 illustrates different sectors' consumption variations due to 5% income increment. Because of different income elasticities, different sectors behave differently versus the income increment. For example, the industrial customers invest a major part of the surplus income on mid-peak consumption.

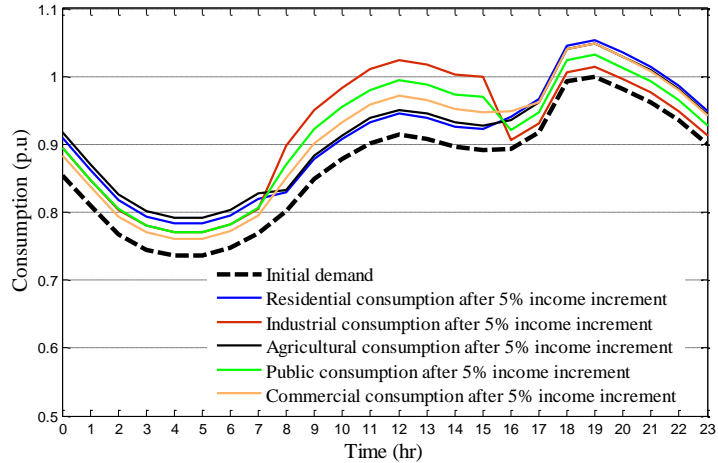

**Fig. 18**: Different sectors' consumption versus 5% income increment.

**Scenario III**

Iranian power market utilizes the single-side auction method to determine the market-clearing price, while the demand side does not bid into the market. In this market, the electricity price has remained regulated to the customers. Each year, the regulator increases the electricity price based on inflation and the customers' average income increment. In this scenario, we want to examine how the consumption will respond if the regulator increases electricity price as much as the income increases. Figure 19 illustrates the Iranian power system demand for different scenarios. From the figure, the system demand does not change if the electricity price increment be equal to the income increment. Moreover, the outcome can be concluded by utilizing (52). It can be concluded to study electricity consumption sensitivity, it is necessary to consider the price and income changes, simultaneously.

Total electricity consumption, budget, and welfare for the different scenarios are compared in Table 12. From the table, the electricity price increment in scenario I leads to load reduction. In this scenario, customers manage the budget through the load reduction, which results in low utility and welfare. The budget is increased equal to 5% in scenario II. The customers spend the surplus budget on consuming more electrical energy. The consumption increment increases the customers' welfare and utility. The consumption will not change in scenario III, when the electricity price grows equal to the budget increment. For this scenario, although the budget is increased, the customers' utility does not change.

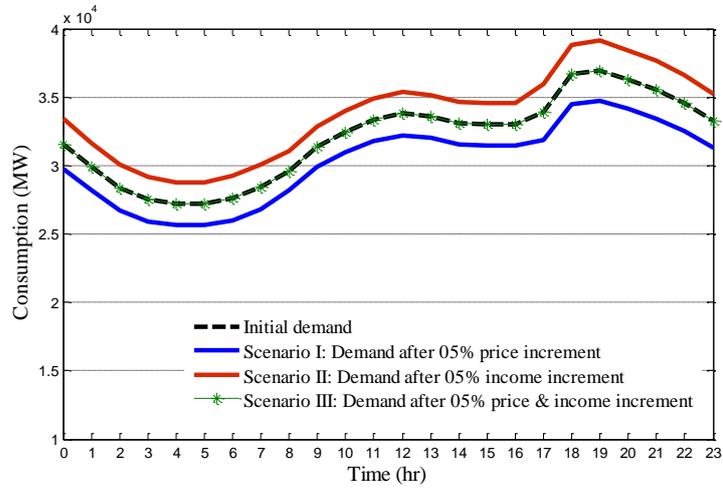

**Fig. 19**: Iranian power system demand considering electricity 5% price and income increment.

**Table 12:** Electricity demand, budget, and welfare considering different Scenarios.

|  | Initial State | Scenario I | Scenario II | Scenario III |
|---|---|---|---|---|
| Demand (MW/day) | 768299 | 727682 | 810123 | 768299 |
| Utiltiy (billion Rial/day) | 766.44 | 730.6 | 808.02 | 766.52 |
| Budget (billion Rial/day) | 468.81 | 468.41 | 492.71 | 492.28 |
| Welfare (billion Rial/day) | 299.63 | 262.19 | 315.31 | 274.24 |

## 6. Conclusion

This paper illustrated that the critical data are embedded in the electricity customers' risk aversion coefficients. Utilizing the concept of the utility function, the coefficients are used to estimate electricity customers' satisfaction due to different consumption levels. Besides, the customers' welfare is calculated based on these coefficients. Finally, a comprehensive load model is designed concerning the electricity price and the customers' risk aversion coefficients. The load model is utilized to estimate the price and income elasticities of electricity demand.

From the results, as the risk aversion coefficient increases, the consumers achieve more satisfaction from the consumption. Also, for the low-risk aversion coefficients, the consumption decreases rapidly with the price increment. Electricity consumption will be complement (substitute) if the risk aversion coefficient is greater (lower) than the unit. For the unit risk aversion coefficient, the cross-price elasticity will be zero, which means the consumption is independent of the other period's demand.

When the electricity price increases, electricity consumers try to manage their load through consumption reduction. The consumption reduction leads to lower utility and welfare. When the customers' income increases, they spend the extra budget on consuming more electricity energy, which increases their satisfaction and welfare. The consumption variations due to the price or income changes are related to the customers' risk aversion coefficients. The results illustrated that the electricity price increment can neutralize the income increment effect. Hence, to study electricity consumption, the price and income variations should be considered, simultaneously.